\documentclass[a4paper,10pt,twoside]{cpc-hepnp}

\usepackage{multicol}
\usepackage{graphicx}
\usepackage{booktabs}
\usepackage{amssymb,bm,mathrsfs,bbm,amscd}
\usepackage[tbtags]{amsmath}
\usepackage{lastpage}
\usepackage{bbding}
\usepackage{pifont}

\usepackage{epsfig}
\usepackage{CJK}
\usepackage{amsmath}
\usepackage{amsfonts}
\usepackage{color}
\usepackage{extarrows}

\def\bc{\begin{center}}

\def\ec{\end{center}}
\def\be{\begin{eqnarray}}
\def\ee{\end{eqnarray}}

\def\om{\omega}

\def\d#1#2{\frac{\displaystyle #1}{\displaystyle #2}}

\def\La{\Lambda}

\begin{document}

\fancyfoot[C]{\small 010201-\thepage}


\title{Quantized  space-time and its influences on some physical problems\thanks{Supported by National Natural Science
Foundation of China (11247261;11175109) }}

\author{%
      Meng-Sen Ma$^{1,2;1)}$\email{mengsenma@gmail.com}%
\quad Hui-Hua Zhao$^{1,2}$%
}

\maketitle

\address{%
$^1$ Department of Physics, Shanxi Datong University, Datong 037009, China\\
$^2$ Institute of theoretical physics, Shanxi Datong University,
Datong 037009, China }

\begin{abstract}
Based on the idea of quantized space-time of Snyder, we derive new
generalized uncertainty principle and new modified density of
states. Accordingly we discuss the influences of the modified
density of states on some physical quantities and laws. In addition
we analyzed the exact solution of the harmonic oscillator in
Snyder's quantized space-time.
\end{abstract}



\footnotetext[0]{\hspace*{-3mm}\raisebox{0.3ex}{$\scriptstyle\copyright$}2013
Chinese Physical Society and the Institute of High Energy Physics
of the Chinese Academy of Sciences and the Institute
of Modern Physics of the Chinese Academy of Sciences and IOP Publishing Ltd}%

\begin{multicols}{2}

\section{Introduction}

Recently a great interest has been devoted to the study of the
generalized uncertainty principle (GUP)
\cite{konishi,maggiore,kempf,kempf1,kempf2,chang,chang1,medved,setare,bang,nasseri,zhao1,bina,said}.
The main consequence of the GUP is the existence of a minimal length
scale of the order of the Planck length, which can be  deduced in
string theory and other theories of quantum gravity
\cite{veneziano,gross,amati,witten,garay,scardigli,adler1,adler2}.
Although at present no this kind of minimal length is probed
experimentally, from the form of the GUP or the general commutation
relation it should exist theoretically. The minimal length can
provide a natural ultraviolet (UV) cut-off. The GUP can also
influence the structure of phase space and modify the usual density
of states  to a different form with a weighted factor. With the
modified density of states one can calculate its effects on
cosmological constant and black body radiation\cite{chang1}
 and  the entanglement entropy of black holes by means of statistical mechanics\cite{zhao}.

As mentioned above the existence of minimum length can be the result
of generalized uncertainty principle. In fact one can also consider
the problem in an opposite direction.
  To deal with the trouble in quantum field theory,
Snyder\cite{snyder} proposed an suggestion that the usual four
dimensional space-time may not be continuous but discrete or
quantized. This means that there is a smallest unit of length in
space-time and the space-time should be noncommutative. Based on the
assumption of existence of a minimal length
 and Lorentz invariance, Snyder introduced some operators for position, momentum and angular momentum and obtained
 a sequence of commutators between them. It shows that the usual commutation relation
  $[\mathcal {\hat X},\mathcal {\hat P}]=i$ and
 $[\mathcal {\hat X},\mathcal {\hat X}]=0$ no more exist. According to the commutators obtained by Snyder we can also derive
a new GUP. Besides one can also obtain the
 modified density of states different from the one obtained from the usual GUP. Thus one can recalculate the influences of the new
 modified density of states on the physical quantities and laws, such as the cosmological constant, black body radiation,
 the entanglement entropy of black holes and so on.
 Particularly the minimal length in Snyder's model give a natural UV cutoff in the
 calculation of entanglement entropy, thus the UV divergence caused by the infinite density near the
horizon can be removed.

 The paper is arranged as follows.  In the next section we first introduce Snyder's quantized space-time model and derive the GUP and modified density of states
. In the third section, we calculate its influences on the
cosmological constant and specially on the Stefan-Boltzmann laws. In
addition we simply analyze the exact solution of the harmonic
oscillator in Snyder's quantized space-time.
 We shall give some concluding remarks in the final section.($c=\hbar=G=k_{B}=1$).

\section{Snyder's quantized space-time and GUP}

Snyder developed the quantized space-time which is invariant under
Lorentz transformation, namely the quadratic form
$-\eta^2=\eta_0^2-\eta_1^2-\eta_2^2-\eta_3^2-\eta_4^2$ should be
invariant under Lorentz transformation. The $\eta_b, (b=0,1,2,3,4)$
are the homogeneous projective coordinates of a real 4-dimensional
space of constant curvature. Thus the space-time operator $\hat
x_\mu, (\mu=0,1,2,3)$ can be defined as

\begin{eqnarray}
\begin{array}{l}
\hat
x_0=ia(\eta_{4}\frac{\partial}{\partial\eta_{0}}+\eta_{0}\frac{\partial}{\partial\eta_{4}});
\quad \hat x_1=ia(\eta_{4}\frac{\partial}{\partial\eta_{1}}-\eta_{1}\frac{\partial}{\partial\eta_{4}}) \\
\hat
x_2=ia(\eta_{4}\frac{\partial}{\partial\eta_{2}}-\eta_{2}\frac{\partial}{\partial\eta_{4}});
 \quad \hat x_3=ia(\eta_{4}\frac{\partial}{\partial\eta_{3}}-\eta_{3}\frac{\partial}{\partial\eta_{4}})
\end{array}
\end{eqnarray}
where $a$ is the minimum length.

In addition there are another two groups of operators:
\begin{eqnarray}
\begin{array}{l}
\hat p_0=\frac{1}{a}(\eta_0/\eta_4);\quad \hat p_1= \frac{1}{a}(\eta_1/\eta_4)\\
\hat p_2=\frac{1}{a}(\eta_2/\eta_4);\quad \hat p_3=
\frac{1}{a}(\eta_3/\eta_4)
\end{array}
\end{eqnarray}
and
\begin{eqnarray}
\begin{array}{l}
\hat
L_1=i(\eta_{3}\frac{\partial}{\partial\eta_{2}}-\eta_{2}\frac{\partial}{\partial\eta_{3}}),\;
\hat
L_2=i(\eta_{1}\frac{\partial}{\partial\eta_{3}}-\eta_{3}\frac{\partial}{\partial\eta_{1}}),\;\\
\hat L_3=i(\eta_{2}\frac{\partial}{\partial\eta_{1}}-\eta_{1}\frac{\partial}{\partial\eta_{3}})\\
\hat
M_1=i(\eta_{0}\frac{\partial}{\partial\eta_{1}}+\eta_{1}\frac{\partial}{\partial\eta_{0}}),\;
\hat
M_2=i(\eta_{0}\frac{\partial}{\partial\eta_{2}}+\eta_{2}\frac{\partial}{\partial\eta_{0}}),\;\\
\hat
M_3=i(\eta_{0}\frac{\partial}{\partial\eta_{3}}+\eta_{3}\frac{\partial}{\partial\eta_{0}})
\end{array}
\end{eqnarray}
We know that Lorentz group has six generators which can be recorded
as $M_{ij}(i,j=1,2,3)$ and $M_{0i}$. The three generators for
rotation $L_i=\frac{1}{2}\epsilon_{ijk}M_{jk}$ and the other three
ones for boost can be described as $M_i=M_{0i}$. It is easy to get
the commutator below:
\begin{eqnarray}
[\hat x_i,\hat x_j]=ia^2 \hat M_{ij}, \quad [\hat x_0, \hat
x_i]=ia^2 \hat M_{0i}
\end{eqnarray}
Obviously if one takes the limit $a \rightarrow0$, the quantized and
noncommutative space-time will turn into the usual continuous and
commutative space-time. In fact what we really care about is the
commutators below:
\begin{eqnarray}\label{com1}
[\hat x_i,\hat p_j]=i(\delta_{ij}+a^2\hat p_i\hat p_j)
\end{eqnarray}
This relation reminds us of the more general commutator from GUP,
which is\cite{kempf1}
\begin{eqnarray}
\label{com2}[\hat x_i,\hat p_j]=i(\delta_{ij}+\lambda
\delta_{ij}\hat p^2+\lambda'\hat p_i\hat p_j)
\end{eqnarray}
One can easily find out that the commutator Eq.(\ref{com1}) from
quantized space-time is a special case of the
 commutators Eq.(\ref{com2}) from GUP with $\lambda=0$ and $\lambda'=a^2$.

In general it is known that for any pair of observables A and B, the
uncertainty relation
\begin{eqnarray}\label{com3}
\bigtriangleup  A\bigtriangleup  B\geq \frac{1}{2}|\overline{[A,
B]}|
\end{eqnarray}
In view of $\bigtriangleup A= A-\bar A$ and $(\bigtriangleup
A)^2=\overline {A^2}-\overline {A}^2$, we can obtain from Eq.
(\ref{com1}) that
\begin{eqnarray}
\bigtriangleup x_i \bigtriangleup p_j \geq
\frac{1}{2}(\delta_{ij}+a^2\bigtriangleup p_i\bigtriangleup
p_j+\gamma)
\end{eqnarray}
where $\gamma$ is positive and dependent on the expectation value of
$p_i$. We can name the GUP above as Snyder's GUP. If considering the
$i=j$ case only, the formula above will turns into
\begin{eqnarray}
\bigtriangleup x_i \bigtriangleup p_i \geq
\frac{1}{2}[1+a^2(\bigtriangleup p_i)^2+\gamma]
\end{eqnarray}
or
\begin{eqnarray}\label{gup}
\bigtriangleup x_i  \geq
\frac{1}{2}\left(\frac{1+\gamma}{\bigtriangleup
p_i}+a^2\bigtriangleup p_i\right)
\end{eqnarray}
which is similar to the frequently used GUP. Obviously when setting
$\gamma=0$, it will give a minimal uncertainty length
$\bigtriangleup x=a$.

According to the usual Heisenberg uncertainty principle, one can
obtain the D dimensional phase space volume
\begin{eqnarray}
d^D\textbf{x}d^D\textbf{p}
\end{eqnarray}
Upon quantization, the corresponding number of quantum states per
momentum space volume is
\begin{eqnarray}
\frac{d^D\textbf{x}d^D\textbf{p}}{(2\pi)^D}
\end{eqnarray}
Considering the commutator  Eq.(\ref{com2}) from GUP, the number of
quantum states changes to\cite{chang}
\begin{eqnarray}\label{com3}
\frac{d^D\textbf{x}d^D\textbf{p}}{(2\pi)^D(1+\lambda
p^2)^{D-1}[1+(\lambda+\lambda')p^2]^{1-\frac{\lambda'}{2(\lambda+\lambda')}}}
\end{eqnarray}
 Thus the number of quantum states for quantized space-time
 should be
 \begin{eqnarray}\label{com4}
\frac{d^D\textbf{x}d^D\textbf{p}}{(2\pi)^D(1+a^2p^2)^{1/2}}
\end{eqnarray}
In fact its counterpart
 \begin{eqnarray}\label{com5}
\frac{d^D\textbf{x}d^D\textbf{p}}{(2\pi)^D(1+\lambda p^2)^D}
\end{eqnarray}
which corresponds to $\lambda'=0$ in Eq.(\ref{com3}) is often
considered by physicists. The difference between the two forms  lies
in the weighted factor, more precisely, the exponent there. The
exponent in Eq.(\ref{com5}) is dimension-dependent, whereas the one
in Eq.(\ref{com4}) is a constant $1/2$. Thus when the two forms of
number of quantum states are used in quantum field theory, one can
deduce that  Eq.(\ref{com5}) can remove the divergence more
effectively, specially, the higher the space dimension is, the
weaker the divergence is. Be that as it may, the Eq.(\ref{com4})
should be employed if the space-time is really discrete as Snyder's
proposition. One can use the formula to recalculate many quantities,
like black body radiation, cosmological constant,et.al.

\section{The influences on some physical problems}
\subsection{The cosmological constant}
According to quantum field theory, the cosmological constant should
be obtained by summing over the zero-point fluctuation energies of
harmonic oscillators, each of which corresponds to a particular
particle momentum state. If the dispersion relation is the usual one
$E^2=p^2+m^2$ (if modified dispersion relation is considered, the
discussion below should be modified correspondently), we assume that
the zero-point energy of each oscillator is of the form
\be
\d{\omega}{2}=\d{1}{2}\sqrt{p^2+m^2}
\ee
Thus the cosmological constant
should be
\be
\La(m)&=&\d{1}{2}\int\d{d^3p}{(1+a^2p^2)^{1/2}}\sqrt{p^2+m^2}\nonumber\\
&=&2\pi\int_{0}^{\infty}\d{p^2dp}{(1+a^2p^2)^{1/2}}\sqrt{p^2+m^2}
\ee
Obviously the integral is power-law divergent. The weighted
factor can only weaken the divergence but cannot cancel it. However,
because of the existence of a minimal length $a$, the momentum
cannot be integrated to infinity. The minimal length means there
should be some $p_{max}$ above which one cannot probe and observe.
Taking the $p_{max}\sim 1/a$, for the $m=0$ case one can obtain a
finite result \be \La=\d{2\pi(2-\sqrt{2})}{3a^4}\sim
\d{2\pi(2-\sqrt{2})}{3}p^4_{max} \ee

\subsection{The Stefan-Boltzmann law}

We discuss the black-body radiation and consider the radiation field
as photon gas. In general the quantum states with momentum from
$p\sim p+dp$ in volume $V$ is
\be
\d{V}{\pi^2}p^2dp=\d{V}{\pi^2}\om^2d\om
\ee
here we have considered
the spin degeneracy of photons and $\varepsilon=\om=p$. The average
quantum number should be
\be
\d{V}{\pi^2}\d{\om^2d\om}{e^{\om/T}-1}
\ee
The internal energy of the photon gas is
\be
U(\om,T)d\om=\d{V}{\pi^2}\d{\om^3d\om}{e^{\om/T}-1}
\ee
which is
Planck formula. Integrating the equation above one can obtain
\be
U&=&\d{V}{\pi^2}\int_0^\infty\d{\om^3d\om}{e^{\om/T}-1}
\xlongequal{x=\om/T}\d{VT^4}{\pi^2}\int_0^\infty\d{x^3dx}{e^x-1}\nonumber\\
&=&\d{\pi^2V}{15}T^4
\ee
which can give the usual Stefan-Boltzmann
law.

Considering the quantized space-time, the density of states is
modified by Eq.(\ref{com4}). Thus the internal energy of the photon
gas should be
\end{multicols}
\be\label{sbl}
U&=&\d{V}{\pi^2}\int_0^\infty\d{\om^3d\om}{(1+a^2\om^2)^{1/2}(e^{\om/T}-1)}
\xlongequal{x=\om/T}\d{VT^4}{\pi^2}\int_0^\infty\d{x^3dx}{(1+a^2T^2x^2)^{1/2}(e^x-1)}\nonumber\\
&=&\d{VT^4}{\pi^2}\int_0^\infty \left[1-\d{1}{2}(aTx)^2+\d{3}{8}(aTx)^4-\d{5}{16}(aTx)^6+...\right]\d{x^3dx}{e^x-1}\nonumber\\
&=&\d{VT^4}{\pi^2}\left[\int_0^\infty\d{x^3dx}{e^x-1}-\d{a^2T^2}{2}\int_0^\infty\d{x^5dx}{e^x-1}+\d{3a^4T^4}{8}\int_0^\infty\d{x^7dx}{e^x-1}-...\right]
\ee
\begin{multicols}{2}
It is know that \be
\int_0^\infty\d{x^{\alpha-1}dx}{e^x-1}=\Gamma(\alpha)\zeta(\alpha)
\ee where $\Gamma(\alpha)$ and $\zeta(\alpha)$ are the Gamma
function and Riemann zeta function respectively. Because the minimal
length $a$ should be very small, the series expansion at $a=0$ is
appropriate. Thus Eq.(\ref{sbl}) can be calculated exactly \be
U&=&\d{VT^4}{\pi^2}\left(\d{\pi^4}{15}-\d{4\pi^6a^2T^2}{63}+\d{3\pi^8a^4T^4}{15}-...\right)\nonumber\\
&=&\d{\pi^2VT^4}{15}\left(1-\d{60\pi^2a^2T^2}{63}+3\pi^4a^4T^4-...\right)
\ee The first term is the usual Stefan-Boltzmann relation and the
latter ones are correction terms. For the usual Stefan-Boltzmann law
we know that there is a maximal frequency, $x=\om_m/T\approx2.82$,
corresponding to the maximal internal energy. According to the
modified expression of internal energy for the photon gas, we can
also find the maximal frequency $\om_{mq}$. Given different values
of $aT$ one can find the maximal values of the modified expression.
The Fig.1 shows that the bigger the value of $aT$ is, the smaller
the values of the maximal frequency and the maximal internal energy
are.
\begin{center}
\includegraphics[scale=0.6]{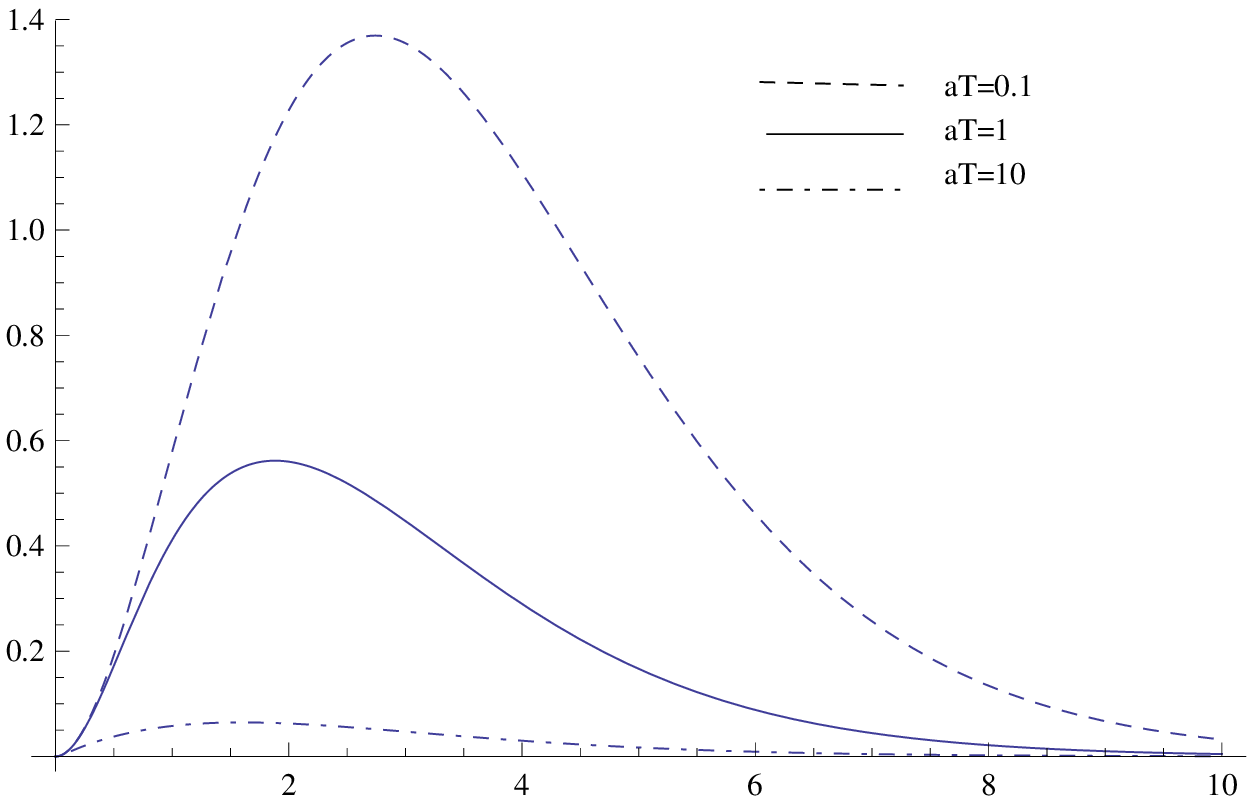}
\figcaption{\label{fig1}  For the case of $aT=0.1$ the maximal value lies
at $x=2.74$ , the case of $aT=1$ at $x=1.88$, and the case of
$aT=10$ at $x=1.60$. }
\end{center}

\subsection{The exact solutions of harmonic oscillator}

Snyder gives the expression for $x_i$ in the momentum
representation. \be\label{xo}
x_i=i(\delta_{ij}+a^2p_i^2\delta_{ij}+a^2p_ip_j)\d{\partial}{\partial
p_j} \ee In the one-dimensional case, it turns into \be
x=i(1+a^2p^2)\d{\partial}{\partial p} \ee which is nearly the same
as the one introduced according to GUP\cite{kempf1,kempf2,chang}, in
which case the position operators
 is $x=i\left[(1+\beta p^2)\d{\partial}{\partial p}+\gamma p\right]$. For the Hamiltonian $H=\d{1}{2}m\om^2x+\d{p^2}{2m}$,
Chang obtained an $\gamma$-independent exact solution\cite{chang}.
Thus referring to the result of Chang, in the Snyder's quantized
space-time model we can deduce that \be
E_n=\om\left[(n+\d{1}{2})\sqrt{1+\d{a^4m^2\om^2}{4}}+(n^2+n+\d{1}{2})\d{a^2m\om}{2}\right]
\ee Once the higher dimensional case is considered, the results must
be different. The general position operator introduced according to
GUP is \be\label{xo1} x_i=i(\delta_{ij}+\beta
p^2\delta_{ij}+\beta'p_ip_j)\d{\partial}{\partial p_j} \ee The main
difference lies at the second term. In Eq.(\ref{xo}) it is $p_i^2$
(no summation here) and in Eq.(\ref{xo1}) it is $p^2$. Thus the two
operators must correspond to different exact solutions in higher
dimensional case. We will leave the question for further
consideration.

\section{Conclusion}
According to the idea of quantized space-time of Snyder, we derive
the generalized uncertainty principle and modified density of
states. The density of states obtained from Snyder's model is
different from the ones from the usual GUP. The weighted factor in
the modified density of states is $1/(1+a^2p^2)^{1/2}$ with a
constant exponent $1/2$, whereas the  one  from usual GUP is
$1/(1+\lambda p^2)^{D}$ with a dimension-dependent exponent $D$.
This difference leads to different modifications to the ordinary
physical problems.

Based on the Snyder's GUP we calculate the cosmological constant.
The
 minimal length gives an natural ultraviolet cutoff, which makes the cosmological constant be proportional to $p_{max}^4$ and finite.
The Stefan-Boltzmann laws in thermodynamics may be also modified
because of the modified density of states. Except the usual $\sim
T^4$ term some correction terms also exist. Considering the modified
Stefan-Boltzmann laws, the rate of black holes radiation will be
influenced and the evolution of the universe should also be
modified. These problems will be discussed in another paper. At last
we discussed the exact solution of harmonic oscillator in
one-dimensional case. The result is nearly the same as the one
obtained in general GUP. When higher-dimensional case is considered
the exact solutions will be different duo to different position
operators. This problem is left for further consideration.

\vspace{-1mm}
\centerline{\rule{80mm}{0.1pt}}
\vspace{2mm}


\end{multicols}

\clearpage


\begin{thebibliography}{90}

\vspace{3mm}

\bibitem{konishi}K. Konishi, G. Paffuti, and P. Provero, \textit{Phys. Lett}. \textbf{B234}, 276(1990).
\bibitem{maggiore}M. Maggiore, \textit{Phys. Lett}. \textbf{B304}, 65(1993); \textit{Phys. Rev}. \textbf{D49}, 5182(1994).
\bibitem{kempf}A. Kempf,  J. Math. Phys. 35, 4483-4496 (1994).
\bibitem{kempf1}A. Kempf, G. Mangano, and R. B. Mann, \textit{Phys. Rev}. \textbf{D52}, 1108(1995).
\bibitem{kempf2}A. Kempf, J. Phys. A 30, 2093(1997).
\bibitem{chang}L. N. Chang, D. Minic, N. Okamura, et al. \textit{Phys. Rev}. \textbf{D65}, 125027(2002).
\bibitem{chang1}L. N. Chang, D. Minic, N. Okamura, et al. \textit{Phys. Rev}. \textbf{D65}, 125028(2002).
\bibitem{medved}A. J. M. Medved and E. C. Vagenas, Phys. Rev. D 70 (2004) 124021 ;
\bibitem{setare}M. R. Setare, Phys. Rev. D 70, 087501(2004);  Int. J. Mod. Phys. A 21, 1325(2006).
\bibitem{bang}J. Y. Bang, and M. S. Berger, \textit{Phys. Rev}. \textbf{D74}, 125012(2006).
\bibitem{nasseri}F. Nasseri, \textit{Phys. Lett}. \textbf{B632}, 151(2006).
\bibitem{zhao1}R. Zhao, and S. L. Zhang, \textit{Phys. Lett}. \textbf{B641}, 208(2006); H. X. Zhao, H. F. Li, S. Q. Hu, R. Zhao,
Commun.Theor.Phys., 48, 465(2007).
\bibitem{bina}A. Bina, S. Jalalzadeh, and A. Moslehi, \textit{Phys. Rev}. \textbf{D81}, 023528(2010).
\bibitem{said}J. L. Said, and K. Z. Adami, \textit{Phys. Rev}. \textbf{D83}, 043008(2011).

\bibitem{veneziano}G. Veneziano,  Europhys. Lett. 2, 199(1986).
\bibitem{gross}D. J. Gross , P. F. Mende, \textit{Nucl. Phys. }\textbf{B303}, 407(1988).
\bibitem{amati}D. Amati, M. Ciafolini, and G. Veneziano, \textit{Phys. Lett}. \textbf{B216}, 41(1989).
\bibitem{witten}E. Witten, Physics today 49, 24(1996).
\bibitem{garay}L. J. Garay, Int. J. Mod. Phys. A10, 145 (1995).
\bibitem{scardigli}F. Scardigli, \textit{Phys. Lett}. \textbf{B452}, 39(1999).
\bibitem{adler1}R. J. Adler, D. I. Santiago, Mod. \textit{Phys. Lett}. \textbf{A14},1371(1999).
\bibitem{adler2}R. J. Adler, P. Chen, and D. I. Santiago, Gen. Rel. Gra., 33,2101(2001).


\bibitem{zhao}R. Zhao, Y. Q. Wu, and L. C. Zhang,  \textit{Class Quantum Grav}. \textbf{20}, 4885(2003).
\bibitem{snyder}H. S. Snyder,  \textit{Phys. Rev}. 71, 38(1947).

\end{thebibliography}
\end{document}